\documentclass{elsart}
\usepackage{psfig}
\usepackage{natbib}
\begin{document}
\runauthor{Polletta, M. and Courvoisier, T.J.-L.}
\begin{frontmatter}
\title{ISOPHOT Observations of Narrow-Line Seyfert 1 Galaxies}
\author[Obs,ISDC]{M. Polletta} and
\author[Obs,ISDC]{T.J.-L. Courvoisier}
\address[Obs]{Geneva Observatory, Ch. des Maillettes 11, CH-1290 Sauverny,
Switzerland}
\address[ISDC]{Integral Science Data Centre, Ch. d'Ecogia 16, CH-1290 Versoix,
Switzerland}
\begin{abstract}
Broad infrared spectra (7--200 $\mu$m) of four NLS1 galaxies, obtained with
the imaging photo-polarimeter (ISOPHOT) on board the Infrared Space
Observatory (ISO), are presented. The infrared luminosities and 
temperatures, opacities and sizes of the emitting dust components are
derived.

A comparison between the observed infrared spectra and the optical emission
line fluxes of a sample of 16 NLS1 galaxies suggests that these objects
suffer different degrees of dust absorption according to the inclination of the
line of sight with respect to the dust distribution.
\end{abstract}
\begin{keyword}
galaxies: Seyfert; galaxies: photometry; infrared: galaxies
\end{keyword}
\end{frontmatter}
%
\section{Introduction}
\label{intro}
The dust properties in NLS1 galaxies may play a key role in explaining
their peculiar properties. Indeed, the relative weakness of low density
forbidden line emission and the absence of broad hydrogen wings in NLS1
galaxies may be understood if most of the ionizing flux is absorbed in the
inner forbidden line region, thus reducing the amount of ionizing radiation
reaching the low density clouds, and absorbing the internal broad line
region (BLR) emission. The presence of dust is indicated by several observations. 
Spectropolarimetry observations suggest that dust scatters the optical
photons in several NLS1 galaxies~\cite{Goodrich89}, and soft X-ray
observations indicate the presence of a dusty warm absorber along the line
of sight to some NLS1 galaxies~\cite{Komossa98}. However, in the
far-infrared (FIR) domain, NLS1s show a high degree of similarity with
normal Seyfert 1 galaxies~\cite{R.Pascual97},\cite{Ulvestad95}, suggesting
that their dust properties are not unique among Seyfert galaxies.
Our purpose is to investigate the dust properties in NLS1 galaxies through
the analysis of their infrared (IR) emission, a direct tracer of
dust~\cite{Barvainis90},\cite{Clavel89}.
%
\section{Spectral energy distributions}
\label{spec_fit}
\begin{figure}[ht] 
      \psfig{figure= 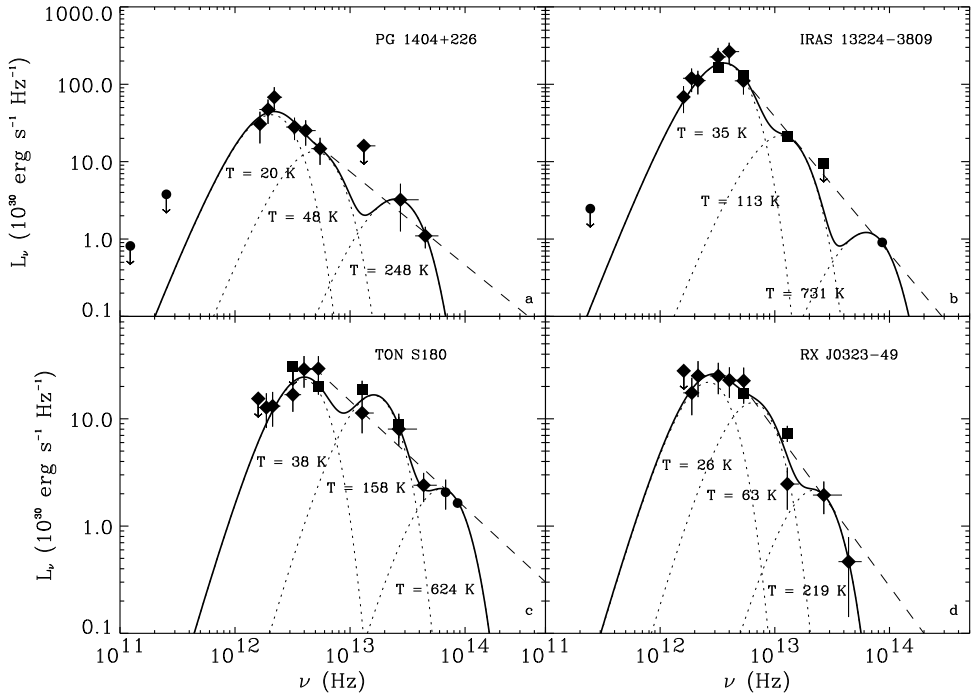,width=13.cm}
      \caption[]{Spectral energy distribution of PG1404+226 (a), IRAS13224$-$3809
   (b), TONS180 (c) and RXJ0323$-$49 (d). Dotted lines represent single grey
   body components at the indicated temperature, and the solid line the
   composite spectrum obtained summing each single component.  The dashed
   line represents the power law obtained fitting data between 3 and 60
   $\mu$m. Arrows indicate 5$\sigma$ upper limits (3$\sigma$ for millimeter
   data). Diamonds represent ISOPHOT data, squares IRAS data, and circles 
   IRAC1, SEST and IRAM data.}
         \label{FigSED}
   \end{figure}

The spectral energy distribution (SED) from near-IR (NIR) to millimetre
wavelengths~\cite{Polletta99} of four NLS1 galaxies (TONS180, RXJ0323$-$49,
IRAS13224$-$3809, and PG1404+226) are analyzed here and shown
in the objects' rest-frames in Fig.~\ref{FigSED}.
%
The grey body radiation model~\cite{Gear88} was used to fit the observed
SEDs. The parameters of the model are the radius $r$ of the projected
source, the dust temperature T, and the dust optical depth
$\tau_\mathrm{d}$.  The observed IR SEDs are smooth, describable by several
grey body components characterized by different temperatures and opacities.
Since the data do not permit us to constrain $\tau_\mathrm{d}$, we fixed it at
two extreme values and then we derived the other parameters of the model.
First, we fixed $\tau_\mathrm{d}$ at one of the highest hydrogen column
density values measured in Seyfert galaxies (1.8$\cdot$10$^{24}$ atoms
cm$^{-2}$), and, secondly, we chose $\tau_\mathrm{d}$ corresponding to the
optical extinction estimated from the Balmer decrement of each
object~\cite{Boller93},~\cite{Miller92},~\cite{Comastri98}, 
~\cite{Grupe96} (see best fitting curves in Fig.~\ref{FigSED}).  In the
case of large opacities, the estimated sizes of the dust emission regions
vary from 0.2 pc for the hottest component (1000 K), to $\sim$500 pc for
the coldest one (20 K), and in the case of low opacities, the minimum
estimated size is 0.3 pc, and the maximum 20 kpc. The comparison between the
temperatures and sizes measured in the two extreme regimes, and the observed
values in Seyfert galaxies suggests that the cold dust is characterized by
low opacity, while the warm/hot dust may be more opaque. As a consequence, the
IR reradiation at long wavelengths (T $\leq$ 60 K) will be isotropic, and
orientation-independent. Conversely, at shorter wavelengths (T $>$ 60
K) the dust may self-absorb its emitted radiation, and the observed emission
may depend on orientation.
%
\section{The role of the dust in NLS1 galaxies}
\label{dust_cont}
\begin{figure}
      \psfig{figure=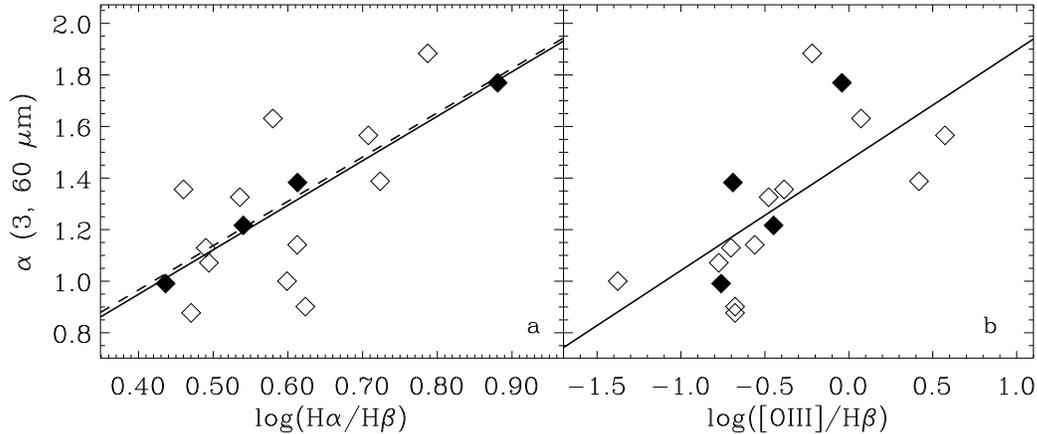,width=14cm}
      \caption[]{Balmer decrement values (a) and line ratios
       [OIII]/H$_{\beta}$ (b) $vs$ the IR spectral index $\alpha_{3.60\mu m}$.
       Filled diamonds represent the 4 NLS1 observed with ISOPHOT and
       empty diamonds the 12 additional NLS1 galaxies. The solid line is
       the best linear fit obtained considering all the 16 NLS1
       galaxies, and the dashed line is the best linear fit obtained
       considering only the 4 NLS1.}
\label{red_IR}
\end{figure}

The steepness of the power law obtained by fitting all the IR data at
wavelengths shorter than 60 $\mu$m, where the effects of opacity become
important, is an indicator of dust extinction. We compared the Balmer
decrement values (H$_{\alpha}$/H$_{\beta}$), indicators of extinction at
optical wavelengths, of the NLS1 galaxies studied here with the slope of
the power law obtained by fitting the observed data between 3 and 60 $\mu$m
(see best-fit power laws in Fig.~\ref{FigSED}). The measured power law
slopes range between 0.99 and 1.77, which corresponds to a range in the visual
extinction of 61 mag, provided that all objects have the same
intrinsic IR power law. The variation in the H$_{\alpha}$/H$_{\beta}$ values
corresponds instead to a range in the visual extinction of only 2.8 mag.
Such a difference indicates that all the emitting dust can not be
responsible for the reddening observed in the Balmer lines. However, the two
parameters, H$_{\alpha}$/H$_{\beta}$, and $\alpha_{3,60\mu m}$, are strongly
correlated (the Spearman's correlation rank is 1.0, with a null associated
probability to obtain such a value from uncorrelated values). The
corresponding values and their linear interpolation are shown in
Fig.~\ref{red_IR}a,  with filled diamonds and a dashed line, respectively. We
selected 12 additional NLS1 galaxies from a sample of 148 objects for
which the Balmer decrement and the IR spectrum were measured. The
observed correlation is still present for the additional 12 NLS1 galaxies
(the Spearman's correlation rank is 0.61, and the probability to obtain such
a value from uncorrelated values is 1.13\%), as shown in
Fig.~\ref{red_IR}a (the solid line represents the best fit line
obtained considering all the 16 NLS1 galaxies which are marked by empty diamonds).
A correlation between $\alpha_{3,60\mu m}$ and the line intensity ratio
[OIII]/H$_{\beta}$ was also discovered (see Fig.~\ref{red_IR}b). The
corresponding Spearman's coefficient rank is 0.78, and the probability to
obtain such a value from uncorrelated values is 0.04\%.
\\
%
%

The steepening of the IR power law and the enhancement of the Balmer
decrement and of the line ratio [OIII]/H$_{\rm \beta}$ can be interpreted
in terms of inclination-dependent obscuration. The central regions of these
objects are probably surrounded by a distribution of dust that becomes
gradually less opaque along some directions. The BLR must be more extended
than the inner hot dust, since its emission is less absorbed than the
``broad" lines. Towards the direction where dust is less opaque, the BLR and
the hottest dust component will be directly visible, while towards the
direction where dust is more opaque, the broad emission lines and the hottest
dust radiation will be partially absorbed. Since the H$_{\beta}$ emission is
produced in both the BLR and in the narrow line region, when the BLR is
obscured the ratio [OIII]/H$_{\beta}$ will be higher than in unobscured
objects.
%
%

\end{document}